\documentclass[runningheads]{llncs}
\usepackage[T1]{fontenc}

\usepackage{graphics}      % standard graphics specifications
\usepackage{graphicx}      % alternative graphics specifications
\usepackage{longtable}     % helps with long table options
\usepackage{url}           % for on-line citations
\usepackage{hyperref}
\usepackage{bm}            % special 'bold-math' package
\usepackage[usenames]{color}
\usepackage[dvipsnames]{xcolor}
\usepackage{cancel}
\usepackage{soul}
\usepackage{amsmath,amssymb}
\usepackage{epstopdf}

\usepackage{ulem}

\newcommand{\steveb}[1]{\textcolor{black}{#1}}

\newcommand{\Th}[1]{\textcolor{black}{#1}}

\begin{document}

\title{Attractive-repulsive challenge in swarmalators with time-dependent speed.}


\author{Steve J. Kongni\inst{1,4} \orcidID{0000-0001-6036-8905} 
\and Thierry Njougouo\inst{2,3,4}\orcidID{0000-0001-7706-7674}
\and Ga\"{e}l R. Simo \inst{4,5} \orcidID{0000-0002-8285-6667}
\and Patrick Louodop\inst{1,4,6,8}\orcidID{0000-0002-2975-2420}
\and Robert Tchitnga\inst{7}
%\orcidID{2222--3333-4444-5555} 
\and Hilda A. Cerdeira\inst{8,9}\orcidID{0000-0003-4805-4668}}
\authorrunning{S.J. Kongni et al.}
\institute{Research Unit Condensed Matter, Electronics and Signal Processing, University of Dschang, %P.O. Box 67 
Dschang, Cameroon.
\and IMT School for Advanced Studies Lucca, Lucca, Italy.
\and Faculty of Computer Science and Namur Institute for Complex Systems (naXys), University of Namur, Namur, Belgium. 
\and MoCLiS Research Group, Dschang, Cameroon.
\and{\textit{Laboratory of Electrotechnics, Automatics and Energy, Higher Technical Teachers, Training College (ENSET) of Ebolowa, University of Ebolowa,Cameroon} }  
%\and ICTP South American Institute for Fundamental Research, S\~ao Paulo State University (UNESP), Instituto de F\'{i}sica Te\'{o}rica, Bloco II, Barra Funda, 01140-070 S\~ao Paulo, Brazil.
\and Potsdam Institute for Climate Impact Research (PIK) Member of the Leibniz Association, % P.O. Box 60 12 03 D-14412 
Potsdam, Germany.
\and Institute of Surface Chemistry and Catalysis, University of Ulm, Albert-Einstein-Allee 47, 89081 Ulm, Germany
\and S\~ao Paulo State University (UNESP), Instituto de F\'{i}sica Te\'{o}rica, S\~ao Paulo, Brazil.
\and Epistemic ME, S\~ao Paulo, Brazil.\\} 


\date{\today}% It is always \today, today,

\maketitle

\begin{abstract}

\Th{We examine a network of entities whose internal and external dynamics are intricately coupled, modeled through the concept of ``swarmalators'' as introduced by O'Keeffe et al. \textcolor{blue}{\cite{o2017oscillators}}. We investigate how the entities' natural velocities impact the network's collective dynamics and path to synchronization. Specifically, we analyze two scenarios: one in which each entity has an individual natural velocity, and another where a group velocity is defined by the average of all velocities. Our findings reveal two distinct forms of phase synchronization---static and rotational---each preceded by a complex state of attractive-repulsive interactions between entities. This interaction phase, which depends sensitively on initial conditions, allows for selective modulation within the network. By adjusting initial parameters, we can isolate specific entities to experience attractive-repulsive interactions distinct from the group, prior to the onset of full synchronization. This nuanced dependency on initial conditions offers valuable insights into the role of natural velocities in tuning synchronization behavior within coupled dynamic networks.
}\\

\end{abstract}

%\pacs{05.45.Xt, 05.45.Jn, 05.45.-a}

%%%%%%%%%%%%%%%%%%%%%%%%%%%%%%%%%%%%%%%%%%%%%%%%%%%%%%%%%%%%%%%%%%%%%%%%%%%%%%
\section{Introduction}
%\noindent
The study, understanding, and synthesis of swarm and oscillator dynamics have recently attracted considerable attention from researchers. This research primarily aims to enhance our understanding of, and to replicate, the behaviors exhibited by various living systems, including birds, fish, frogs, and social insects. Over the last decade, a new class of systems known as mobile systems has emerged, characterized by the coupling of internal and spatial dynamics. These systems, referred to as swarmalators by O'Keeffe et al. \cite{o2017oscillators}, merge the dynamics of swarming with those of oscillators. Numerous studies in this domain have highlighted phenomena such as external sinusoidal excitation \cite{lizarraga2020synchronization}, phase transitions \cite{kongni2023phase}, and various other interactions \cite{o2018ring,o2019review,hong2021coupling,sar2022swarmalators,barcis2020sandsbots,ghosh2023antiphase,hong2023swarmalators,ceron2023,sar2023pinning}.

Building on the two-dimensional model of swarmalators, Sar et al. developed a one-dimensional version, demonstrating that random pinning can induce chaotic behavior within the system \cite{sar2023pinning}. Furthermore, researchers have explored this system in various configurations within one-dimensional and two-dimensional spaces. However, much of the existing research has concentrated on the swarmalator model primarily through pairwise interactions. Recently, the focus has shifted toward incorporating higher-order interactions within swarmalator systems to better approximate the dynamics of specific real-world scenarios \cite{anwar2024collective}.

Importantly, to the best of our knowledge, all existing research on swarmalator dynamics has primarily aimed at identifying new dynamics and transitions toward synchronization, while often assuming that particles do not possess self-propulsion velocities. The attractive and repulsive interactions described in the model by O'Keeffe et al. \cite{o2017oscillators} play a critical role in the spatial mobility of entities. However, incorporating a self-propulsion velocity for each entity would more accurately reflect the behavior observed in many living systems, capturing their individual responses to external factors and adaptations to group movement.

In this work, we incorporate a non-zero self-propulsion speed to investigate its impact on the transition toward a synchronized state. Following this introduction, we will present the studied model in Section 2. In Section 3, we will analyze the numerical results, focusing first on the effects of individual self-propulsion speed, followed by the effects of group self-propulsion speed. We will conclude with a summary of our findings.

%%%%%%%%%%%%%%%%%%%%%%%%%%%%%%% body %%%%%%%%%%%%%%%%%%%%%%%%%%%%%%%%%%%%%%%%%
%%%%%%%%%%%%%%%%%%%%%%%%%%%%%%%%%%%%%%%%%%%%%%%%%%%%%%%%%%%%%%%%%%%%%%%%%%%%%%

\section{Model}

\Th{Let us consider a model, which has spurred extensive research into swarmalator dynamics, encompassing diverse configurations, coupling mechanisms, and emergent behaviors \textcolor{blue}{\cite{o2017oscillators,lizarraga2020synchronization,o2018ring,hong2021coupling,sar2022swarmalators}}. This system is described by Eqs. \ref{e1} and \ref{e2}, which capture the core interactions underlying these complex dynamics.}

\begin{equation}\label{e1}
{{\dot r}_i} = {v_i} + \frac{1}{N}\sum\limits_{j \ne i}^N {{F_{att}}\left( {{r_{ij}}} \right)W\left( {{\theta _{ij}}} \right) - {F_{rep}}\left( {{r_{ij}}} \right)},
\end{equation}

\begin{equation}\label{e2}
{{\dot \theta }_i} = {w_i}{\mkern 1mu}  + \frac{K}{N}\sum\limits_{j \ne i}^N {{H_{att}}\left( {{\theta _{ij}}} \right)G\left( {{r_{ij}}} \right)}, 
\end{equation}\\
\Th{with $i, j = 1, \dots, N$, where $N$ denotes the total number of swarmalators; $\theta_i$ represents the phase of the internal dynamics of each entity, and $r_i = (x_i, y_i) \in \mathbb{R}^2$ denotes the spatial coordinate of the $i^{\text{th}}$ swarmalator. The parameters $v_i$ and $w_i$ correspond to the velocity and natural frequency of each entity, respectively.
}
The attractive and repulsive interactions between entities in the network are represented by the three explicit functions $F_{att}$, $H_{att}$, and $F_{rep}$, respectively \textcolor{blue}{\cite{o2017oscillators,lizarraga2020synchronization}}. 
\Th{The influence of the internal dynamics on the oscillators' movement is represented by the functions \( W \) and \( G \), which also capture the reciprocal effect of movement on internal dynamics.}
\Th{The model described by Eqs.\ref{e1} and \ref{e2} can be reformulated as follows:
\begin{equation}\label{ee1}
{{{\dot r}_i} = \alpha V(t) + \frac{1}{N}\sum\limits_{j \ne i}^N {\underbrace {\frac{{{r_{ij}}}}{{\left\| {{r_j}_i} \right\|}}\left( {A + J\cos \left( {{\theta _{ij}}} \right)} \right)}_{Attractive{\kern 1pt} {\kern 1pt} term} - \underbrace {B\frac{{{r_{ij}}}}{{{{\left\| {{r_j}_i} \right\|}^2}}}}_{{\rm{Repulsive}}{\kern 1pt} {\kern 1pt} {\rm{term}}}} },
\end{equation}
\begin{equation}\label{ee2}
{{\dot \theta }_i} = {w_i} + \frac{K}{N}\sum\limits_{j \ne i}^N {\frac{{\sin \left( {{\theta _{ij}}} \right)}}{{\left\| {{r_j}_i} \right\|}}}.
\end{equation}
Here, \( r_{ij} = r_j - r_i \) and \( \theta_{ij} = \theta_j - \theta_i \). The constant \( \alpha \) is chosen within \([0,1]\); \( A \) and \( B \) are set to 1; \( K \) represents the phase coupling strength, and the interaction between spatial and phase dynamics is modulated by \( A + J \cos \left( \theta_{j} - \theta_{i} \right) \).   \( J \) measures the influence of phase similarity on spatial attraction.}
\Th{Previous studies on swarmalators often simplify their models by assuming both the natural velocity \( v_i \) and natural frequency \( w_i \) are zero. In contrast, we will set \( w_i = 0 \) while allowing \( v_i \neq 0 \) in our analysis.
}
\Th{Assuming zero intrinsic speed fails to accurately describe the dynamics of many living systems. For instance, in a flock of birds or a school of fish, the intrinsic speed of individuals may vary based on their internal states or external factors, such as the search for food or the presence of a predator. In these scenarios, a member may adjust its speed to pursue food or evade threats, leading to changes in direction, potential shifts in leadership, and variations in the intrinsic speeds of the group members.\\
Building on this example, we propose incorporating the effect of each entity’s individual speed as it returns to the group after an external disturbance. For simplicity, we model each entity’s speed as dependent on its speed from the previous moment (see Eq.\ref{vitp}), representing a memory state of its intrinsic speed.\\
Furthermore, interactions among entities within their collective dynamics have enabled us to define an additional form of speed, the ``group speed'' \( v_g \) (see Eq.\ref{vitg}). This group speed characterizes each entity’s tendency to return to the collective movement pattern established prior to an external disturbance, facilitating coordinated movement and minimizing collision risk.
}

\Th{The two hypotheses formulated above allow us to examine the influence of a group speed \( v_g \) and an individual speed distribution \( v_p \), as defined by Eqs.\ref{vitg} and \ref{vitp}.
\begin{itemize}
\item Delayed group velocity
\begin{equation}\label{vitg}
V(t) = {v_g} = {\left\langle {{{\dot r}_i}(t - \delta t)} \right\rangle _{i = 1...N}}
\end{equation}
\item Delayed individual velocity
\begin{equation}\label{vitp}
V(t) = {v_p} = {{\dot r}_i}(t - \delta t)
\end{equation}
\end{itemize}
\(\left\langle {...} \right\rangle\) represents the time average, and \(\delta t\) denotes the constant time delay, set equal to the integration step \((\delta t = dt = 0.1)\) for simplicity. Initially, all entities begin with zero velocity.\\
We will then analyze how these different choices of natural speeds affect the overall dynamics of the swarmalator network, as well as their potential influence on the network’s progression toward phase synchronization.
}

\section{Numerical results}
\Th{To analyze our model’s dynamics, we simulated a network of 50 nodes using the fourth-order Runge-Kutta method with a time step of \( dt = 0.1 \). Initial spatial positions is randomly assigned from a uniform distribution over \([-1, 1]\), and internal phases is initialized within \([- \pi, \pi]\). Each solution is computed over \(10^6\) iterations. In this section, we examine the effects of both individual and group speed distributions on the network dynamics and further investigate how the choice of initial conditions influences the balance between attractive and repulsive interactions.
}

\subsection{Effect of group velocity to the route to phase synchronized state.}
\begin{figure*}[!h]
\centering
\includegraphics[width=10cm, height=10cm]{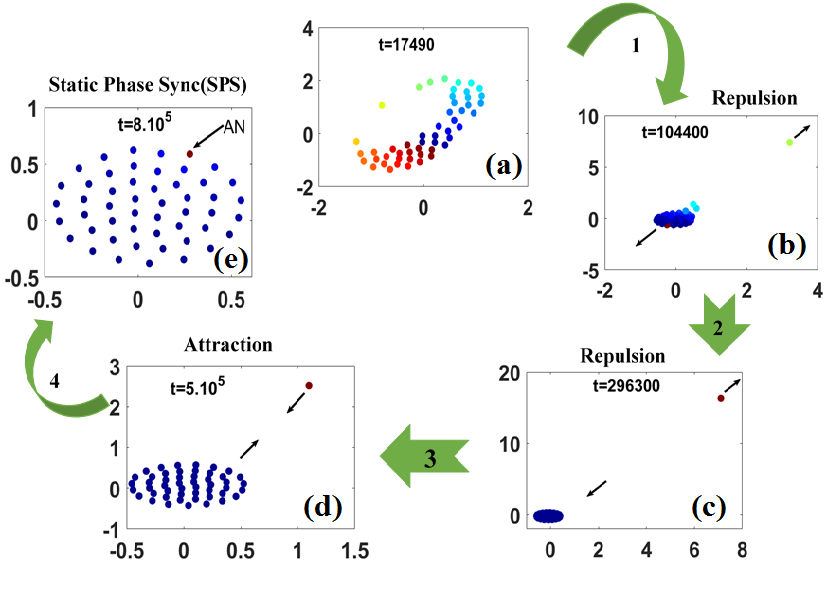}
\caption{Snapshots of states showing the evolution toward synchronization throught the attractive-repulsive challenge with the existence of active node (AN) when the delayed group velocity is applied. (a) Satic wing state; (b,c) repulsion between active node (AN) and the static sync group; (d) attraction between AN with the static sync group; (e) Static phase sync (SPS). Plotted for $J=1, K=0.0004$ and $\alpha=0.3$. The black arrow indicate the sens of displacement of both active node (AN) and the group (see movie 1 in the supplementary material \textcolor{blue}{\cite{supp}}).}
\label{snapshotS}
\end{figure*}

\Th{Let us consider in this initial case that the natural velocity at time \( t \) is uniform and corresponds to the average of velocities of all elements at time \( t - \delta t \), as described by Eq.\ref{vitg}. The dynamics of the system are governed by Eqs.\ref{ee1} and \ref{ee2}.\\
Based on this assumption, Fig.\ref{snapshotS} presents the transition from the static wing state to a static phase-synchronized state. The figure illustrates this progression through snapshots, emphasizing the repulsive and attractive interactions that drive one agent to behave as a solitary wave or ``soliton''. In these snapshots, we observe various transitional states preceding static phase synchronization. These transitions reveal that synchronization is achieved through a balance of competing forces: repulsive interactions that foster the formation of an active solitary node (AN) and attractive interactions that pull this node back toward the group, resulting in stable phase synchronization.
}

\begin{figure}[!h]
\centering
\includegraphics[width=7cm, height=5cm]{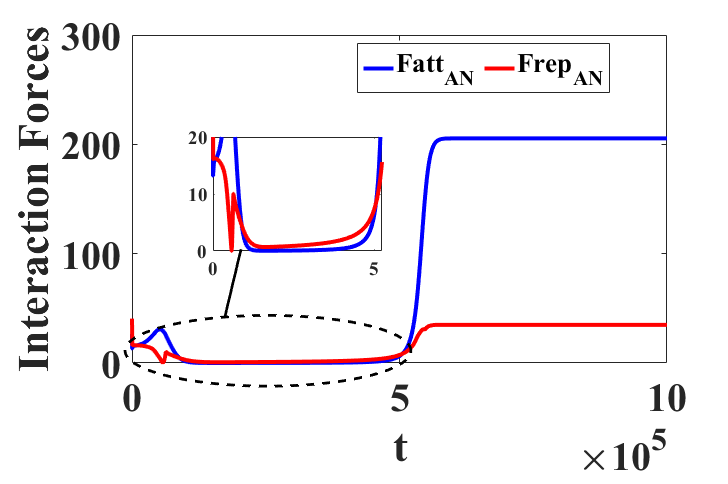}
\caption{Time evolution of the mean of the interactions forces suggested by the active node (AN) when the delayed group velocity is applied. The red color indicate the repulsive interaction while the blue one represent the attractive interaction between the active node and the others elements. Plotted for $J=1, K=0.0004$ and $\alpha=0.3$. }
\label{forcesimpS}
\end{figure}

\Th{Fig.\ref{snapshotS}(a) illustrates the creation of a solitary node at time \( t = 17490 \), driven by the repulsive interaction. When this repulsive force dominates over the attractive interaction, the solitary node moves away from the rest of the group, as seen in the transition from stages 1 to 2 (i.e., Figs.\ref{snapshotS}(b) and \ref{snapshotS}(c)). Stage 3, highlighted in Fig.\ref{snapshotS}(d), demonstrates the return of the solitary node to the group. This stage is characterized by the predominance of the attractive interaction over the repulsive one between the active node and the other elements in the network (see Fig.\ref{forcesimpS}).
}

\begin{figure}[!h]
\centering
\includegraphics[width=7cm, height=5cm]{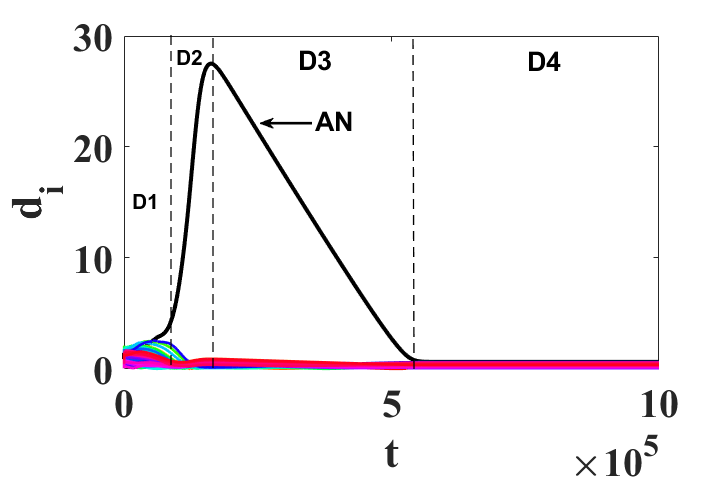}
\caption{Time evolution of the mean radius of all the nodes when the delayed group velocity is applied. The black curve represents the active node (AN). Plotted for $J=1, K=0.0004$ and $\alpha=0.3$}
\label{radiusS}
\end{figure}

To better illustrate the challenge of attractive and repulsive interactions, we represent in Fig.\ref{radiusS} the relative distance of mutual interaction, \( d_i \) (see Eq.\ref{dist}), between the elements of the network. This figure highlights four distinct regions (D1 to D4) corresponding to stages 1 to 4 of Fig.\ref{snapshotS}.

\begin{equation}\label{dist}
d_i(t) = \sqrt{x_i^2(t) + y_i^2(t)} 
\end{equation}

At the initial moment in region D1, the elements are distributed randomly, resulting in a relative distance \( d_i \) whose average value approaches zero. During this time, the attractive and repulsive interactions largely compensate for each other (\( F_{rep} \simeq F_{att} \)), with a slight dominance of the attractive interaction occurring after \( t = 28970 \) (\( F_{att} \ge F_{rep} \)), as indicated by the blue curve in Fig.~\ref{forcesimpS}.

In region D2 (stage 2 in Fig.\ref{snapshotS}(b,c)), we observe the onset of dominance by the repulsive interaction force of the active element, represented by the red curve in Fig.\ref{forcesimpS}, where \( F_{rep} > F_{att} \). This shift is characterized by an increase in the relative distance \( d_i \) of the active element from the rest of the group (see Fig.\ref{radiusS}). The change in concavity of the relative distance observed in region D3 suggests an imminent shift in dominance between the two interaction forces, similar to what we noted in D2. Specifically, the repulsive force continues to prevail in D3, maintaining the relationship \( F_{rep} > F_{att} \), as illustrated in the zoomed-in view of Fig.\ref{forcesimpS}. However, approaching region D4, we witness a tendency for the active node (AN) to be attracted toward the rest of the group, as depicted in Fig.\ref{snapshotS}(d). This indicates a transition to a dominance of the attractive force over the repulsive force, where \( F_{att} > F_{rep} \).

\begin{figure*}[!h]
\centering
\includegraphics[width=6cm, height=4.5cm]{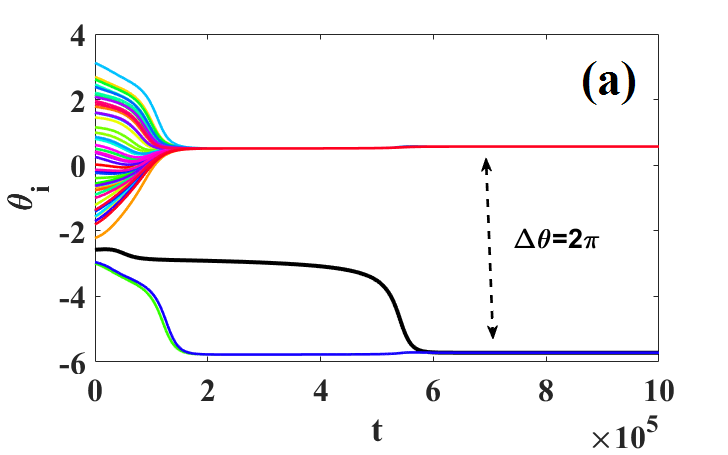}
\includegraphics[width=6cm, height=4.5cm]{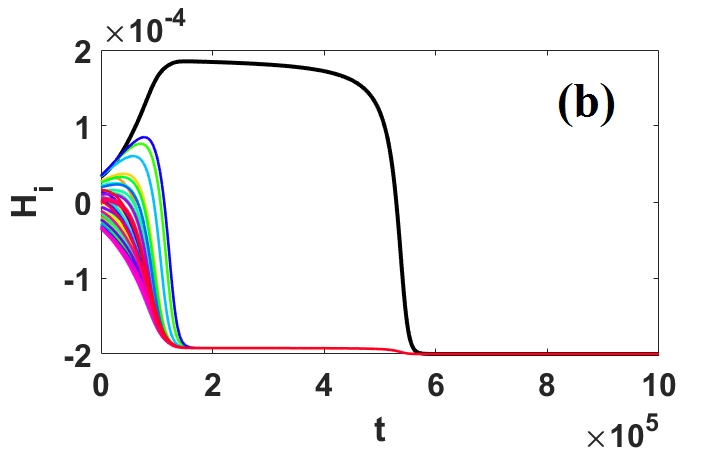}
\caption{Time evolution showing transition to static phase synchronization. (a) The internal phase dynamic $\theta_i$ of each element. (b) The interaction energy $H_i$ of each element. Plotted for $J=1, K=0.0004$ and $\alpha=0.3$.}
\label{FreqenergyS}
\end{figure*}

In addition to the evolutions illustrated in Figs.\ref{snapshotS} and \ref{radiusS}, it is important to highlight that during the transition toward static phase synchronization, the active node demonstrates distinct internal phase dynamics across regions D1 to D3 (see Fig.\ref{FreqenergyS}(a)). Notably, the entire group maintains a consistent phase difference of \(2\pi\) between regions D2 and D3, which is indicative of phase synchronization. However, the active node only achieves a phase difference of either zero or \(2\pi\) starting from region D4.
Conversely, the evaluation of the phase interaction energy \( H_i \) \cite{hong2018active} between the entities reveals a direct relationship with the relative distance of interaction \( d_i \) between the active node and the group. Specifically, as the distance \( d_i \) increases, the interaction energy \( H_i \) also increases, and conversely, it decreases as \( d_i \) decreases. Notably, the interaction energy stabilizes once synchronization is achieved among the internal phases of the entities, as illustrated in Fig.\ref{FreqenergyS}(b).

\Th{The increase in phase interaction energy \( H_i \) with the relative distance \( d_i \) between the active node and the group can be attributed to several factors. As the distance increases, the interaction energy often rises due to the dominance of repulsive forces over attractive ones, resulting in a loss of cohesion and stability within the system. This greater separation leads to increased phase differences, which disrupts synchronization and raises the energetic costs associated with rejoining the group. Additionally, the effective strength of interactions typically diminishes with distance, creating energy barriers that the active node must overcome to synchronize again. Overall, these dynamics reflect the underlying physical principles governing the interactions within the system, where both individual and collective behaviors significantly influence the energy landscape.
}

\subsection{Distributed velocity effect on the transition to phase synchronization.}

\begin{figure}[ht!]
\centering
\includegraphics[width=6cm, height=4.5cm]{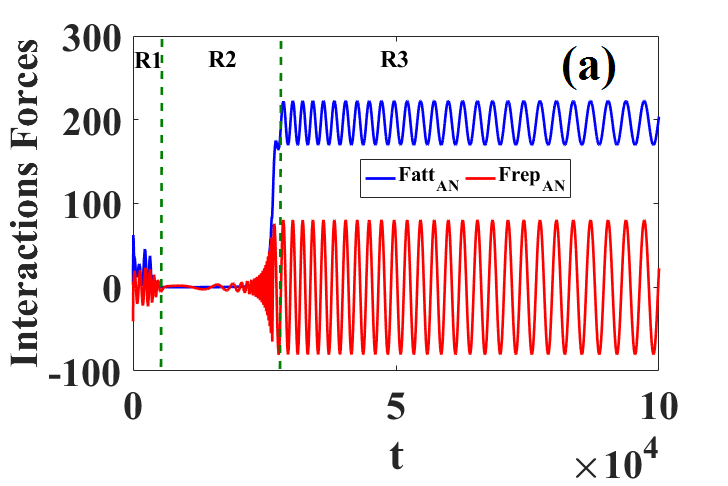}
\includegraphics[width=6cm, height=4.5cm]{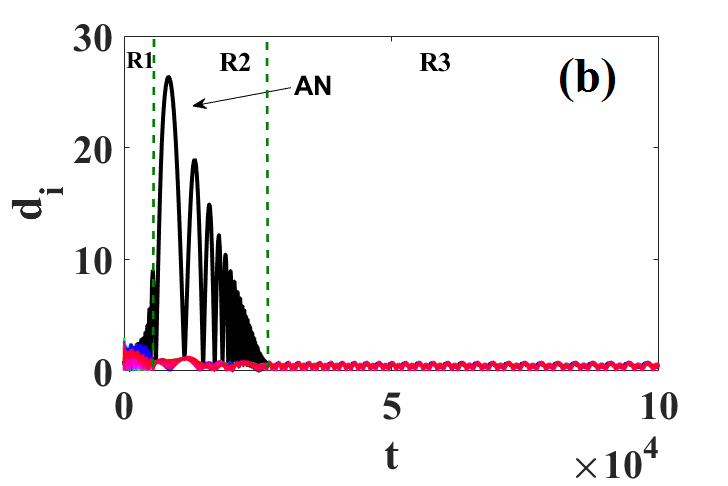}
\caption{Time evolution showing the transition to a static phase synchronization when the delayed individual velocity is applied. (a) Interactions forces, the red color indicates the repulsive interaction while the blue one represents the attractive interaction between the active node and the others elements; (b) The mean radius of the nodes (the black curve is for the active node(AN)). Plotted for $J=1, K=0.0008$ and $\alpha=1$.}
\label{radiusforcesimpR}
\end{figure}

\steveb{Let us now examine the impact of a delayed group speed distribution on the transition to synchronization. Given that the function \( V(t) \) in Eq.\ref{ee1} is defined by Eq.\ref{vitg}, we find that, similar to the effects of delayed individual speeds, this delayed distribution significantly influences not only the transition to synchronization but also the type of synchronization achieved.
\begin{figure*}[ht!]
\centering
\includegraphics[width=10cm, height=12cm]{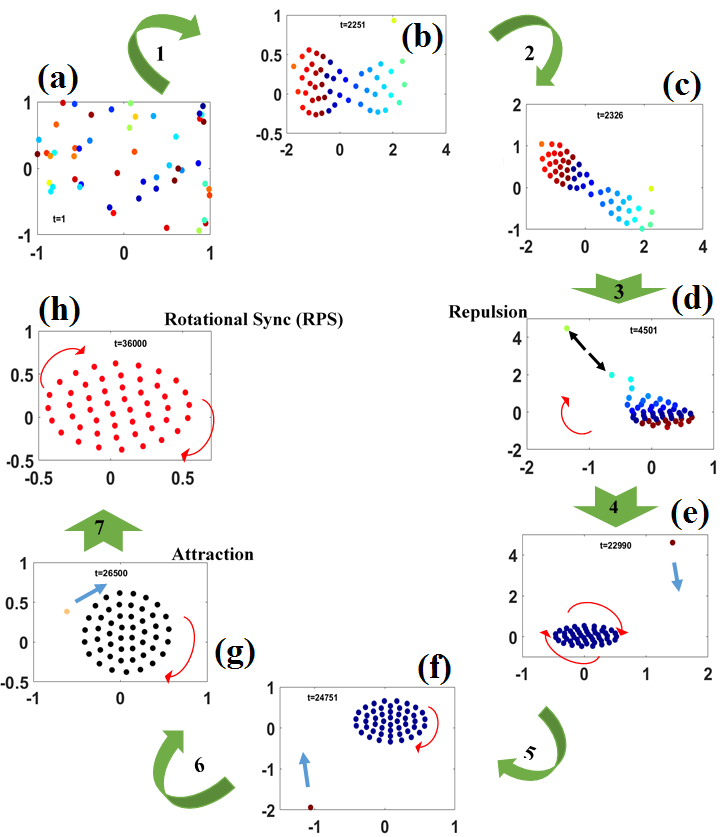}
\caption{Snapshots of states showing the evolution toward synchronization throught the attractive-repulsive challenge with the existence of active node (AN) when the delayed individual velocity is applied. (a) Initial distibution of entities;(b)Repulsion with creation of active node(AN); (c,d) Repulsion between active node (AN) and the dynamic sync group; (e,f,g) Attraction between AN with the dynamic sync group; (h) Rotational phase sync (RPS). Plotted for $J=1, K=0.0008$ and $\alpha=1$, the bleu arrow indicate the sens on displacement of the active node (AN) while the red arrow indicate the sens of rotation of the group (see movie 2 in the supplementary material \textcolor{blue}{\cite{supp}}).}
\label{snapshotR}
\end{figure*}
The snapshots in Fig.\ref{snapshotR} show the different transitional states leading to the synchronization state, where we observe a rotational behavior among the synchronized entities. Thus, by generating the initial conditions as seen in Fig.\ref{snapshotR}(a), we witness the formation of a compact group and a solitary or active node (AN), characteristic of a dominant repulsive interaction (Fig.\ref{snapshotR}(b)). In Figs.\ref{snapshotR}(c) and (d), we observe a repulsive effect between the active node and the group of entities (shown by the opposite black arrow). This repulsion is accompanied by the dominance of the repulsive interaction force over the attractive one ($F_{rep} > F_{att}$), as seen in Fig.\ref{radiusforcesimpR}(a) within region R2. 
}

This dominance is also accompanied by an increase in the interaction distance of the active node relative to the other elements in region R2 of Fig.\ref{radiusforcesimpR}(b), followed by a decrease in this distance as it approaches region R3. Unlike in the case of delayed group speed, the return of the active node here depends not only on a dominant attractive interaction ($F_{\text{att}} > F_{\text{rep}}$) in region R3 but also on the presence of a constant spatial phase difference, $\Delta\phi_{ij} = |\phi_j - \phi_i|$, between the active node and the oscillating group.
\steveb{This search for a constant phase difference can be observed in the continuous movement of the active node in the opposite direction to the rotating group (indicate by the blue arrow in Fig.\ref{snapshotR}(e) and (f)). However, in Fig.\ref{snapshotR}(g), we observe an attraction and penetration of the active node into the rest of the group, with a tangential orientation to the constant rotational movement, leading to the achievement of rotating phase synchronization (RPS) as shown in Fig.\ref{snapshotR}(h).}

We can conclude that achieving a synchronized state, whether under delayed individual speeds or delayed group speeds, requires navigating a simultaneous tension between attractive and repulsive interactions.

\subsection{Initial condition dependence on the existence of attractive-repulsive challenge.}

The synchronization transitions observed above underscore the significant impact of non-zero speed distribution on swarmalator dynamics. However, these transitions appear to be highly sensitive to the choice of initial conditions. Specifically, for a given set of initial conditions, we found that the occurrence of synchronization transitions—and the resulting balance of attractive and repulsive interactions—depends strongly on the initial phase configuration of the entities.

Alternatively, it has been observed that modifying the initial phase condition of the active node alone can successfully adjust its behavior. In other words, this approach enables effective propagation of the phenomenon to the desired element. However, changes to the initial spatial conditions have no impact on the occurrence or progression of this transition toward synchronization.

\steveb{To extend our observation, we applied perturbations $\epsilon_1$ and $\epsilon_2$ to the initial spatial conditions and internal phases, as given in Eqs.\ref{cipert1} and \ref{cipert2}.}   

\begin{equation}\label{cipert1}
{r_i} = {r_i}(0) + {\varepsilon _1}
\end{equation}
\begin{equation}\label{cipert2}
{\theta _i} = {\theta _i}(0) + {\varepsilon _2}
\end{equation}

\begin{figure}[!h]
\centering
\includegraphics[width=8cm, height=5cm]{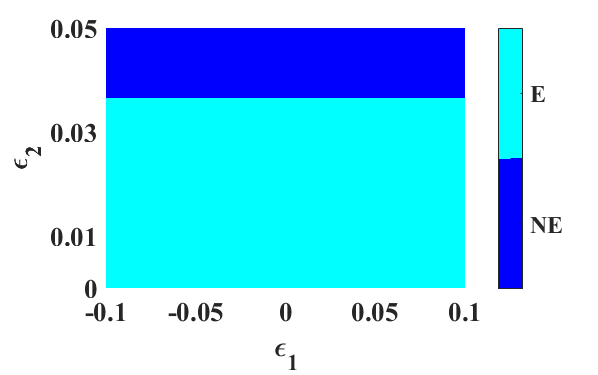}
\caption{Domain of existence (E) and non-existence (NE) of a transition to synchronization with attractive-repulsive challenge. Plotted for $J=1; K=0.0004$ and $\alpha=0.3$.}
\label{2DIC}
\end{figure}

By simultaneously varying \(\epsilon_1\) and \(\epsilon_2\) in the context of delayed group speed, we find that specific conditions can reveal the presence or absence of an attractive-repulsive dynamic between the active node and the network as a whole. Fig.\ref{2DIC} illustrates the domains of existence (E) or non-existence (NE) of synchronization transitions under this dynamic, represented in cyan and blue, respectively. Our observations indicate that perturbations in the spatial variable have minimal influence on the presence of interaction conflicts. Specifically, applying perturbations within the range \({\varepsilon_1} \in \left[ -0.1, 0.1 \right]\) does not alter the synchronization transition. However, perturbations in the internal phase variable \(\epsilon_2\) have a notable impact: as \(\epsilon_2\) increases from 0 to 0.05, it promotes a synchronization transition featuring an attractive-repulsive dynamic, provided that \(\epsilon_2 \leq 0.04\). Beyond this threshold, when \(\epsilon_2 > 0.04\), this transition ceases to exist.

\section{Conclusion}
In this study, we investigated the transition to synchronization in a swarmalator system influenced by non-zero individual speeds, an area that has received less attention in previous research. We focused on how these non-zero speeds affect synchronization dynamics, addressing two distinct formulations.
First, we examined the scenario where the speed of an element at time \(t\) is dependent on the average speed at time \(t - \delta t\), as defined by Eq.\ref{vitg}. Our findings indicate that the transition to synchronization is characterized by a simultaneous emergence of attractive and repulsive interactions. This dynamic results in an increase in the interaction energy of the active node (AN) and culminates in a static phase synchronization state.
In our second approach, we explored the dependency of an entity's speed at the present time \(t\) on its speed at the previous time \(t - \delta t\), as described by Eq.\ref{vitp}. This formulation introduces a memory component to the behavior of the node. Here, we identified a novel state of dynamic synchronization termed Rotational Phase Synchronization (RPS), which is preceded by a transition that also highlights the attractive-repulsive challenge.
Furthermore, we emphasized the significance of initial conditions on the type of transition to synchronization, establishing a domain of existence or non-existence for this behavior. Investigating how initial conditions can influence the number of entities exhibiting this synchronization could provide deeper insights into the dynamics of living systems, such as those observed in flocks of birds or schools of fish. This line of inquiry presents a promising avenue for future research, as it may enhance our understanding of collective behaviors in biological systems.

\begin{credits}
\subsubsection{\ackname} PL and HAC thank ICTP-SAIFR and FAPESP grant 2021/14335-0 for partial support. TN thanks the ``Reconstruction, Resilience and Recovery of Socio-Economic Networks'' RECON-NET EP\_FAIR\_005 - PE0000013 ``FAIR'' - PNRR M4C2 Investment 1.3, financed by the European Union – NextGenerationEU for the partial support. PL thanks the support of the Deutscher Akademischer Austausch Dienst (DAAD) at the Potsdam Institute for Climate Impact Research (PIK) under the Grant Number (91897150).
SJK, TN, PL and MoCLiS research group thank ICTP for the equipment, donated under the letter of donation Trieste $12^{th}$ August 2021. 
.

\subsubsection{\discintname}
The authors have no competing interests to declare that are
relevant to the content of this article.
\end{credits}

% % % % % % % % % % % % % % % References % % % % % % % % % % % %

\nocite{*}
\bibliographystyle{unsrt}%splncs04}
\bibliography{references}

\end{document}

\begin{center}

\bf{SUPPLEMENTARY MATERIAL}

\bf{Attractive-repulsive challenge in swarmalators with time-dependent speed.}
\end{center}

\begin{center}
Steve J. Kongni, Thierry Njougouo, Ga\"{e}l R. Simo, Patrick Louodop, Robert Tchitnga and Hilda A. Cerdeira \\
\end{center}

In this supplemental file, we give the description of the contents of each movie.\\ \\ 

\section*{Description of the movies}\\

\textbf{Supplementary Movie 1:}  Movie 1 shows the effect of group velocity to the route to static phase synchronized state (SS). Each state of the movie is associated with the different snapshots of the Fig.1 in the paper. \\\\
\textbf{Supplementary Movie 2:} Movie 2 highlight the influence of the distributed velocity on the transition to the synchronization. This transition leads to the Rotational phase synchronization (RPS). Each state of the movie is associated with the different snapshots of the Fig.5 in the paper.\\